\documentclass[twocolumn]{aastex63}
\usepackage[T1]{fontenc}
\usepackage[normalem]{ulem}
\usepackage[utf8]{inputenc}

\shorttitle{Direct imaging search for companions on DSHARP disks}
\shortauthors{Jorquera et al.}

\begin{document}

\title{A search for companions via direct imaging in the DSHARP planet-forming disks}

\author{Sebasti\'an Jorquera}
\affiliation{Departamento de Astronom\'ia, Universidad de Chile, Camino El Observatorio 1515, Las Condes, Santiago, Chile}
\author{Laura M. P\'erez}
\affiliation{Departamento de Astronom\'ia, Universidad de Chile, Camino El Observatorio 1515, Las Condes, Santiago, Chile}
\author{Gaël Chauvin}
\affiliation{Departamento de Astronom\'ia, Universidad de Chile, Camino El Observatorio 1515, Las Condes, Santiago, Chile}
\affiliation{Unidad Mixta Internacional Franco-Chilena de Astronom\'ia, CNRS/INSU UMI 3386}
\affiliation{Universit\'e Grenoble Alpes, CNRS, IPAG, 38000 Grenoble, France}
\author{Myriam Benisty}
\affiliation{Departamento de Astronom\'ia, Universidad de Chile, Camino El Observatorio 1515, Las Condes, Santiago, Chile}
\affiliation{Unidad Mixta Internacional Franco-Chilena de Astronom\'ia, CNRS/INSU UMI 3386}
\affiliation{Universit\'e Grenoble Alpes, CNRS, IPAG, 38000 Grenoble, France}
\author{Zhaohuan Zhu}
\affiliation{Department of Physics and Astronomy, University of Nevada, Las Vegas, 4505 S. Maryland Pkwy, Las Vegas, NV, 89154, USA}
\author{Andrea Isella}
\affiliation{Department of Physics and Astronomy, Rice University, 6100 Main Street, Houston, TX 77005, United States of America}
\author{Jane Huang}
\affiliation{Center for Astrophysics \textbar\ Harvard \& Smithsonian, 60 Garden Street, Cambridge, MA 02138, United States of America}
\affiliation{NHFP Sagan Fellow, Department of Astronomy, University of Michigan, 323 West Hall, 1085 S. University Avenue, Ann Arbor, MI 48109, USA}
\author{Luca Ricci}
\affiliation{Department of Physics and Astronomy, California State University Northridge, 18111 Nordhoff Street, Northridge, CA 91330, USA}
\author{Sean M. Andrews}
\affiliation{Center for Astrophysics \textbar\ Harvard \& Smithsonian, 60 Garden Street, Cambridge, MA 02138, United States of America}
\author{Shangjia Zhang}
\affiliation{Department of Physics and Astronomy, University of Nevada, Las Vegas, 4505 S. Maryland Pkwy, Las Vegas, NV, 89154, USA}
\author{John Carpenter}
\affiliation{Joint ALMA Observatory, Avenida Alonso de C\'ordova 3107, Vitacura, Santiago, Chile}
\author{Nicol\'as T. Kurtovic}
\affiliation{Departamento de Astronom\'ia, Universidad de Chile, Camino El Observatorio 1515, Las Condes, Santiago, Chile}
\author{Tilman Birnstiel}
\affiliation{University Observatory, Faculty of Physics, Ludwig-Maximilians-Universit\"at M\"unchen, Scheinerstr.  1, 81679 Munich, Germany}
\affiliation{Exzellenzcluster ORIGINS, Boltzmannstr. 2, D-85748 Garching, Germany}


\begin{abstract}
The "Disk Substructures at High Angular Resolution Project" (DSHARP) has revealed an abundance and ubiquity of rings and gaps over a large sample of young planet-forming disks, which are hypothesised to be induced by the presence of forming planets. In this context, we present the first attempt to directly image these young companions for 10 of the DSHARP disks, by using NaCo/VLT high contrast observations in L'-band instrument and angular differential imaging techniques. We report the detection of a point-like source candidate at 1.1" (174.9 au) for RU Lup, and at 0.42" (55 AU) for Elias 24. In the case of RU Lup, the proper motion of the candidate is consistent with a stationary background contaminant, based on the astrometry derived from our observations and available archival data. For Elias 24 the point-like source candidate is located in one of the disk gaps at 55 AU. Assuming it is a planetary companion, our analysis suggest a mass ranging from $0.5 M_J$ up to $5 M_J$, depending on the presence of a circumplanetary disk and its contribution to the luminosity of the system. However, no clear confirmation is obtained at this stage, and follow-up observations are mandatory to verify if the proposed source is physical, comoving with the stellar host, and associated with a young massive planet sculpting the gap observed at 55\,AU. For all the remaining systems, the lack of detections suggests the presence of planetary companions with masses lower than $5M_J$, based on our derived mass detection limits. This is consistent with predictions of both hydrodynamical simulations and kinematical signatures on the disk, and allows us to set upper limits on the presence of massive planets in these young disks.
\vspace{1cm}
\end{abstract}



\section{Introduction} \label{sec:intro}
Since the first confirmed detection of an exoplanet \citep{1992Natur.355..145W}, almost three decades ago, the search for these objects have revealed a large amount of planetary systems architectures, showing a great diversity in orbital properties (misalignment, period, eccentricity), multiplicity, and mass distribution \citep{2015ARA&A..53..409W}. Multiple factors are believed to influence the final configuration of planetary systems, such as the different mechanisms of planetary formation (e.g., core accretion, \citeauthor{1996Icar..124...62P} \citeyear{1996Icar..124...62P}; gravitational instability, \citeauthor{1978M&P....18....5C} \citeyear{1978M&P....18....5C}, \citeauthor{1997Sci...276.1836B} \citeyear{1997Sci...276.1836B}), the initial conditions of the environments where planets are formed \citep{2014prpl.conf..691B, 2018haex.bookE.143M}, and the dynamical interactions that these systems go through \citep{2016A&A...591A..45P, 2018A&A...618A..18R}. It is then essential to understand the role of each of these elements on the final configuration of the planet population.\par
In this scenario, protoplanetary disks become a key component to understand the formation and evolution of planetary systems, as it is thought that their configuration is closely linked to when and where planets are formed in the disk. Initially believed to be smooth during early stages of evolution, high resolution imaging of protoplanetary disks have revealed the presence of numerous substructures, such as rings/gaps \citep{2017A&A...600A..72F,2018ApJ...869L..41A, 2018MNRAS.475.5296D,2018ApJ...869L..42H,2018ApJ...869...17L}, or spirals \citep{2015ApJ...808L...3A, 2016Sci...353.1519P,2017ApJ...840...32T,2017AJ....154...33A,2018ApJ...869L..43H,2019ApJ...872..122M}, across a wide range of ages \citep[from $<1$ Myr and upwards of 10 Myr, e.g.][]{2017A&A...597A..42B, 2018ApJ...869L..50P, 2018ApJ...857...18S}. The origin of annular rings and gaps, which is the most commonly observed substructure, is still unclear, as many different physical mechanisms can create the observed features. For example, dust accumulation in snowline regions \citep{2012ApJ...752..106O,2015IAUGA..2256118Z, 2017ApJ...845...68P}, or zonal flows due to the magnetorotational instability \citep{2009ApJ...697.1269J, 2014ApJ...784...15S}, may be able to create ringed substructure, but the most commonly assumed explanation is that depleted rings in a protoplanetary disk are induced by  young planetary companions forming inside the disk \citep{2004A&A...425L...9P,2010A&A...518A..16F,2012A&A...545A..81P,2018ApJ...864L..26B}.\par
Multiple studies have been carried out, aiming to show that the existence of annular substructures can be explained with the presence of one or multiple planetary companions embedded into the protoplanetary disk, considering for example the orbital evolution of the planet \citep{2019ApJ...886...62L} and the correlation between the disk and the companion properties \citep{2018ApJ...869L..47Z,2018ApJ...864L..26B}. Although these works put strong constraints on the physical properties of both the disks and the possible companions, which are in agreement with the formation of the observed dark gaps, they have also revealed that the origin of these structures can be explained with different configurations of planetary masses and disks properties (e.g., viscosity, grain size distribution, disk thermodynamics, etc.). So without a prior, independent estimation of these properties, it is not possible to properly conclude whether these gaps are, in fact, evidence of the presence of a planetary companion or not, as it is almost always possible to find a combination of disk and planet properties that could explain the observed gaps. One solution for this problem is to directly observe and characterize the possible companions, to compare with the planetary and disks properties derived from these models. \par
Of the different techniques for planet detection, direct imaging has already proven to be an extremely useful tool for the detection and characterization of possible companions in the presence of transition and debris  disks (HR 8799, \citeauthor{2008Sci...322.1348M} \citeyear{2008Sci...322.1348M}; $\beta$ Pictoris, \citeauthor{2009A&A...493L..21L} \citeyear{2009A&A...493L..21L}; HD 95086, \citeauthor{2013ApJ...779L..26R}\citeyear{2013ApJ...779L..26R}; HD 206893, \citeauthor{2017A&A...597L...2M} \citeyear{2017A&A...597L...2M}; PDS 70, \citeauthor{2018A&A...617A..44K} \citeyear{2018A&A...617A..44K}). This technique offers the unique opportunity to observe the formation of giant planets within their birth environment, study the planet-disk architecture and interactions, and access the orbital parameters together with the luminosity and atmospheric properties of the planets. However, one of the drawbacks is that the planet mass estimation is only accessible through the use of uncalibrated mass-luminosity relation that depends on the initial conditions of formation \citep[the so called cold, warm, and hot-start models,][]{2007ApJ...655..541M}. Extending the search for companions via direct imaging to young protoplanetary disks that exhibit annular substructures, will not only help to shed light into the origin of these structures, but to also provide strong evidence on the different stages of planetary evolution, depending on the age of the observed disks.\par
In this context, the Atacama Large sub-Millimeter Array (ALMA) Large Program titled "Disks Substructures at High Angular Resolution Program" \citep[DSHARP, ][]{2018ApJ...869L..41A} provides a unique sample of young circumstellar disks where to search for planetary companions, as the 20 young systems from the $\rho$-Ophiuchus \citep{2008hsf2.book..351W}, Lupus \citep{2008hsf2.book..295C} and Upper Scorpius \citep{2008hsf2.book..235P} star forming regions have been recently imaged at high spatial resolution at a wavelength of 1.3\,mm. The interest in this sample arises from two key properties: the ubiquity of annular substructures in the objects from the DSHARP sample  \citep{2018ApJ...869L..42H}, and the relatively young age of the systems ($\sim1$ Myr on average), making it ideal to both study the planet-disk interactions that could produce these features and the properties of giant planets in formation or recently formed.\par
In this work, we present our attempt to directly image companions in the disks from the DSHARP sample, using L' ($\lambda_0 = 3.8 \mu\textrm{m}$, $\Delta\lambda = 0.62\mu\textrm{m}$) observations from the Very Large Telescope (VLT) NAOS-CONICA (NaCo) instrument. The paper is structured as it follows: in Section \ref{sec:obs} we present the selection criteria for the observed targets, as well as the observation strategy. We then proceed to describe the data reduction in Section \ref{sec:processing}. In Section \ref{sec:Results} we present the main results, followed by the corresponding discussion in Section \ref{sect:Discussion}. Finally, we conclude with a summary of our main results in Section \ref{sect:Conclusions}.

\begin{longrotatetable}
\label{table:sample}
\begin{deluxetable*}{llccccrrccc}
\tablecaption{NaCo Sample: Star properties}
\tablewidth{700pt}
\tabletypesize{\scriptsize}
\tablehead{
\colhead{Name} & \colhead{Region} & 
\colhead{2MASS} & \colhead{d} & 
\colhead{SpT} & \colhead{log $T_{eff}$} & 
\colhead{log $L_*$} & \colhead{log $M_{*}$} & 
\colhead{log $t_{*}$} & \colhead{log $\dot{M}_*$}& \colhead{L'} \\ 
\colhead{} & \colhead{} & \colhead{Designation} & \colhead{(pc)} & 
\colhead{} & \colhead{(K)} & \colhead{($L_\odot$)} &
\colhead{($M_\odot$)} & \colhead{(yr)} & \colhead{($M_\odot yr^{-1}$)} & \colhead{(mag)} \\
\colhead{(1)} & \colhead{(2)} & \colhead{(3)} & \colhead{(4)} & \colhead{(5)} & \colhead{(6)} & \colhead{(7)} & \colhead{(8)} & \colhead{(9)} & \colhead{(10)} & \colhead{(11)}
}
\startdata
HT Lup & Lup I & J15451286-3417305 & $154\pm2$ & K2 & $3.69\pm0.02$ & $0.74\pm0.20$ & $0.23^{+0.26}_{-0.13}$ & $5.9\pm0.3$ & < -8.4 & $5.8\pm0.12^{(1)}$\\
GW Lup & Lup I & J15464473-3430354 & $155\pm3$ & M1.5 & $3.56\pm0.02$ & $-0.48\pm0.20$ & $-0.34^{+0.10}_{-0.17}$ & $6.3\pm0.4$ & $-9.0\pm0.4$ & $8.4 \pm 0.02^{(2)}$\\
RU Lup & Lup II & J15564230-3749154 & $159\pm3$ & K7 & $3.61\pm0.02$ & $0.16\pm0.20$ & $-0.20^{+0.12}_{-0.11}$ & $5.7\pm0.4$ & $-7.9\pm0.4$ & $6.0 \pm 0.12^{(1)}$\\
Sz 114 & Lup III & J16090185-3905124 & $162\pm3$ & M5 & $3.50\pm0.01$ & $0.69\pm0.20$ & $-0.76^{+0.08}_{-0.07}$ & $6.0^{+0.1}_{-0.8}$ & $-9.1\pm0.3$ & $8.7\pm 0.02^{(2)}$\\
Sz 129 & Lup IV & J15591647-4157102 & $161\pm3$ & K7 & $3.61\pm0.02$ & $-0.36\pm0.20$ & $-0.08^{+0.03}_{-0.15}$ & $6.6\pm0.4$ & $-8.3\pm0.3$ & $7.8\pm 0.03^{(2)}$\\
HD 143006 & Upper Sco & J15583692–2257153 & $165\pm5$ & G7 & $3.75\pm0.02$ & $0.58\pm0.15$ & $0.25^{+0.05}_{-0.08}$ & $6.6\pm0.3$ & $-8.1\pm0.4$ & $5.8\pm 0.12^{1)}$\\
AS 205 & Upper Sco & J16113134-1838259 & $128\pm2$ & K5 & $3.63\pm0.03$ & $0.33\pm0.15$ & $-0.06^{+0.07}_{-0.05}$ & $5.8\pm0.3$ & $-7.4\pm0.4$ & $4.6\pm 0.22^{(2)}$\\
Elias 24 & Oph L1688 & J16262407–2416134 & $136\pm3$ & K5 & $3.63\pm0.03$ & $0.78\pm0.20$ & $-0.11^{+0.16}_{-0.08}$ & $5.3\pm0.4$ & $-6.4\pm0.5$ & $6.6\pm 0.08^{(2)}$\\
WaOph 6 & Oph N 3a & J16484562-1416359 & $123\pm2$ & K6 & $3.62\pm0.03$ & $0.46\pm0.20$ & $-0.17^{+0.17}_{-0.09}$ & $5.5\pm0.5$ & $-6.6\pm0.5$ & $6.2\pm 0.08^{(2)}$\\
AS 209 & Oph N 3a & J16491530–1422087 & $121 \pm 2$  & K5 & $3.63\pm0.03$ & $0.15\pm0.20$ & $-0.08^{+0.11}_{-0.14}$ & $6.0\pm0.4$ & $-7.3\pm0.5$ & $6.4\pm 0.07^{(2)}$
\enddata
\tablecomments{Values extracted from \cite{2018ApJ...869L..41A}. The columns are ordered as follows: Column 1: target name. Column 2: Associated star-forming region. Columnn 3: 2MASS designation. Column 4: distance. Column 5: Spectral type from the literature. Column 6: Effective temperature from the literature. Column 7: Stellar luminosities from the literature, scaled according to the appropriate distance in column 4. Column 8: Stellar masses. Column 9: Stellar ages. Column 10: Accretion rates inferred from accretion luminosities. Column 11: L' magnitudes (magnitude references: (1)\cite{2017yCat.2346....0B}, (2)\cite{2014yCat.2328....0C}).}
\end{deluxetable*}
\end{longrotatetable}

\section{Observations} \label{sec:obs}


The NaCo instrument at VLT is composed of the NAOS Adaptative Optics (AO) system and the near-infrared imager and detector CONICA. Given the limiting sensing magnitude ($K<9-10$\,mag) of the infrared wavefront sensor of NAOS, only 16 systems from the original sample could be selected for the program. Due to the constraints imposed for the observations (sidereal time constraints to observe close to meridian and avoid the zenith), together with the decommisioning of NaCo in October 2019, only 10 of the possible 16 systems were observed. Table \ref{table:sample} compiles the observed sample, together with their stellar hosts properties.

Observations were carried out over more than a year, starting on June 16th, 2018, and ending on  September 29th, 2019.  Given the red colours of our targets, the infrared wavefront sensor was used with the JHK dichroic. The L27 camera was used for data acquisition, which provides a sampling of $\simeq$ 27.1 mas/pixel, together with the broadband L' filter ($\lambda_0 = 3.8 \mu\textrm{m}$, $\Delta\lambda = 0.62\mu\textrm{m}$). 

Pupil-tracking mode, necessary for the use of Angular Differential Imaging (ADI), was used to reduce instrumental speckles that limit the detection performances at inner angles. To optimize frame selection as well as sky and instrumental background removal, we use NaCo cube mode with a window field of view of $512\times520$ pixels (FoV of $\approx 14"\times14"$), a short integration time of 0.2 seconds per frame, and a dithering pattern of five offset positions.

Each observation is composed of a shallow starting and ending sequence with unsaturated exposure using the neutral density filter ND\_Long, with a transmission of $\sim 1.75\%$ for L'. This sequence provides an estimation of the stellar point spread function (PSF) and serves as a photometric calibrator. The central sequence is composed of deep saturated images without the neutral density filter, following the five dithering pattern over a total observing time of 2 hours to maximize the FoV rotation.

In the case of the object Elias 24, images were obtained during two epochs. For the second set of observations, to relax the observation constraints at the meridian passage and secure the re-observation of this system, we opted for a referential differential imaging (RDI) approach. For this, we obtained consecutive saturated science images for Elias 24 and the reference star 2MASS J17354481-2413439 \citep[L' = 6.388 $\pm$ 0.079, ][]{2003yCat.2246....0C} with similar parallactic angle variation. This allows us to use the PSF of the reference star as a model for the PSF of Elias 24 during the RDI reduction process.

The details of the observing setup for each specific target are reported in Appendix A.
\section{Data processing} \label{sec:processing}

All the observations were processed using the IPAG-ADI pipeline \citep{2012A&A...542A..41C}. As a first step, data reduction (flat-fielding, bad pixels, and sky removal) was performed on all the available cubes of each target. Sub-frames of 128 $\times$ 128 pixels (FoV of $ 3.5"\times3.5"$) were extracted to reduce computation time. These frames were then re-centered following a Moffat fitting on the non-saturated part of the stellar PSF wing. The last step was to remove the bad frames (poorly saturated, overly extended PSF) to obtain a final master cube containing all the reduced frames, together with their parallactic angles.

For the ADI subtraction, various flavours of ADI algorithms were applied, namely classical ADI (cADI), smart ADI (sADI), radial ADI (rADI), all described in \cite{2012A&A...542A..41C}, as well as Locally Optimized Combination of Images \citep[LOCI, ][]{2007ApJ...660..770L}, Principal Component Analysis \citep[PCA, ][]{2012ApJ...755L..28S} and Angular Differential Optimal Exoplanet Detection Algorithm \citep[ANDROMEDA, ][]{2015A&A...582A..89C}. The use of multiple ADI techniques allows for comparison and consistency on the results obtained. For sADI and LOCI, we followed a similar configuration as \cite{2012A&A...542A..41C}, with a $FWHM=4.5$ pixels and a separation criteria of $0.75 \times FWHM$ at the companion separation. For the PCA method, observations were reduced using three different number of modes ($k=1,5,20$) with optimization areas of 5, 10, 40 and 60 pixels. No separation criteria was applied in this case.
For the RDI processing of Elias 24, data reduction was carried out using the PCA method as a reference used to model and substract the PSF from the images of Elias 24.

For the platescale and True North calibration used to derive the relative astrometry of the candidates, we considered the solution of the contemporaneous ISPY survey over 2018 and 2019 \citep{2020A&A...635A.162L}: pixel scale of $27.193\pm0.059$\,mas, and True North of $-0.568\pm0.115$\,deg.\footnote{Note that \cite{2020A&A...635A.162L} refer to the True North correction in Table\,C.2 in Appendix.}

\section{Results}\label{sec:Results}
\begin{figure*}[ht!]
\centering
\includegraphics[width=16cm,trim=0cm 0.5cm 0.4cm 1.2cm, clip]{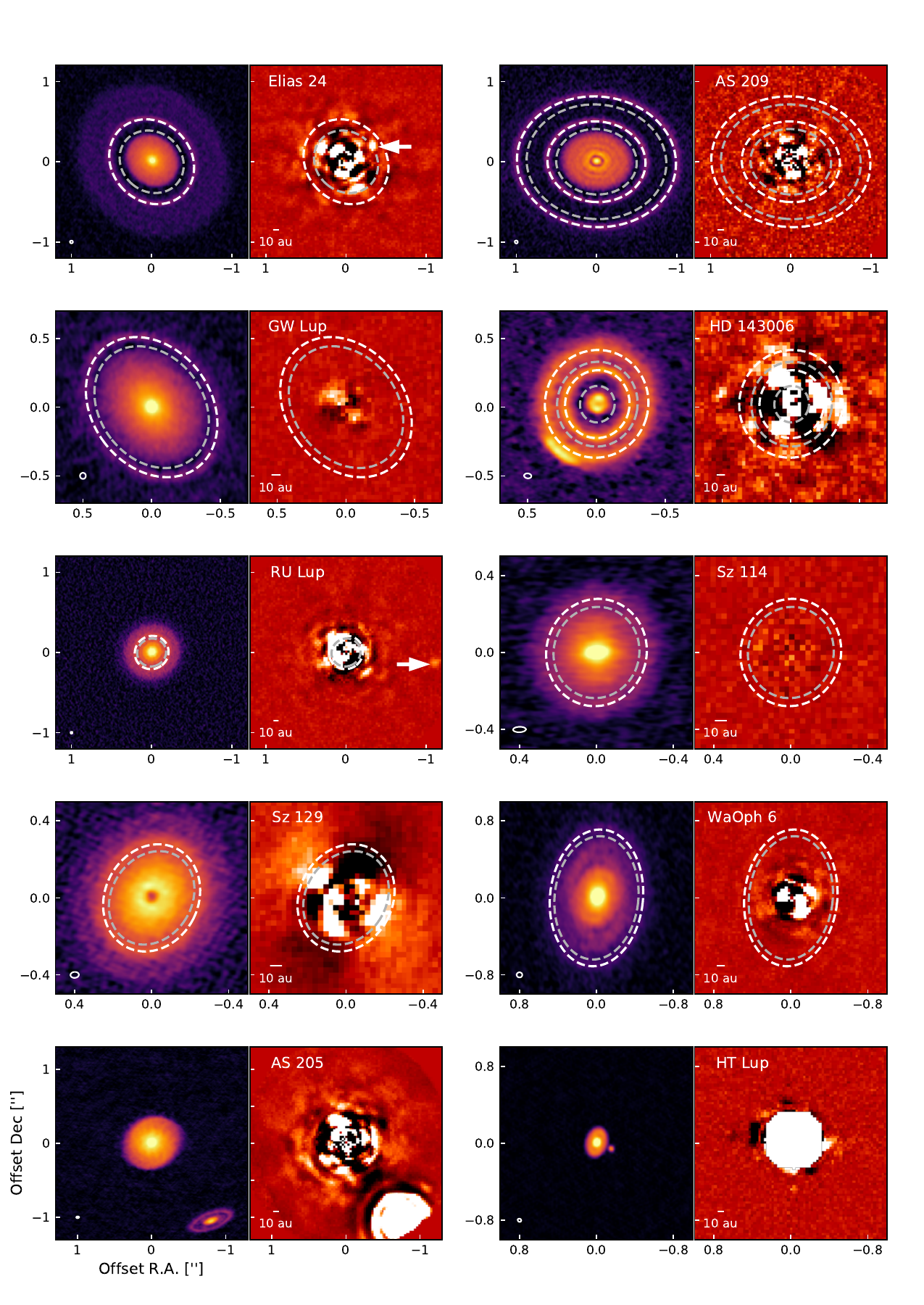}
\caption{ALMA and NaCo  gallery of all observed targets. ALMA images (left panels) are from the DSHARP survey, NaCo images (right panels) are the results from the Classical ADI reduction. White (grey) dashed lines mark the bright rings (dark gaps) of each disk, as derived by \cite{2018ApJ...869L..42H}. Point-like sources are marked by a white arrow when recovered after the ADI reduction. A 10 AU scalebar is shown in the lower left corner of each cADI panel, the beam size of the ALMA images is shown in the lower left corner of each ALMA panel. In the case of HT Lup, variation of the parallactic angle was too low for proper cADI processing. The result from the PCA processing is presented instead.}
\label{fig:sample}
\end{figure*}

We obtained images of all targets using all ADI algorithms discussed in Section \ref{sec:processing}; for simplicity we only present in Fig.\,\ref{fig:sample} the images from the cADI reduction method. We note these cADI images are consistent with the resulting images from all the reduction algorithms discussed above. Figure \ref{fig:sample} also compares the NaCO/cADI images with the respective DSHARP images at 1.3\,mm from \citet{2018ApJ...869L..41A,2018ApJ...869L..42H,2018ApJ...869L..43H,2018ApJ...869L..44K,2018ApJ...869L..48G,2018ApJ...869L..50P}, where for each source we highlight the location of its corresponding bright rings (white dashed lines) and dark gaps (grey dashed lines). In the majority of cases, the ADI reduction was limited to separations $\ge0.1-0.2"$, excluding the possibility of detections in the innermost regions of the systems (below $20-40$ AU). 



\begin{figure}[ht!]
\epsscale{1.2}
\plotone{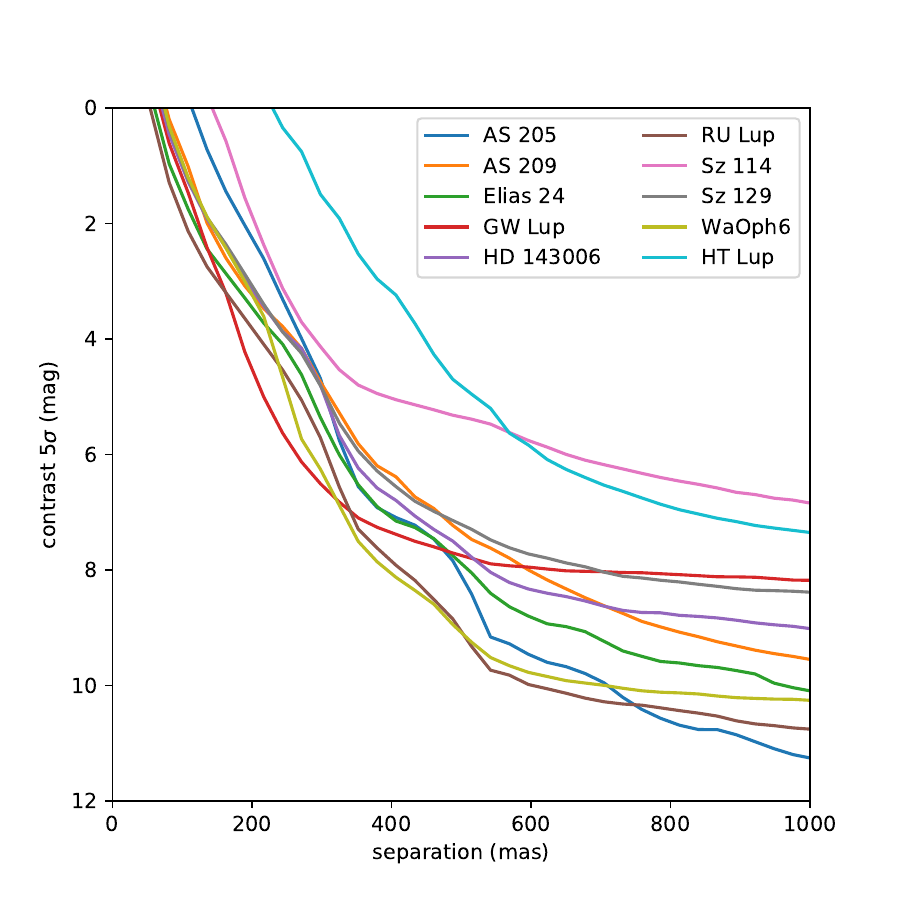}
\caption{5$\sigma$ contrast using the sADI algorithm as a function of separation for all the targets in our sample.}
\label{fig:contrast}
\end{figure}

\subsection{Detection Limits}\label{subsec:limits}

Regardless of the presence or lack of a detection of a possible planetary signal in the objects analyzed, it is possible to extract significant information regarding the  limit at which a detection will be feasible, in both magnitude and mass, of each individual target.

Pixel-to-pixel noise maps of each ADI reduction were estimated within a sliding box of $1.5\times1.5$\,\textit{FWHM} in the NaCo field of view. To correct for the flux loss related to the ADI processing, fake planets were regularly injected (every 10 pixels at 3 different position angles, with a flux corresponding to 100\,ADU) on the original data cubes before PSF subtraction. The cubes were then reprocessed using the different ADI algorithms, allowing us to estimate the flux loss \textbf{by computing the azimuthal average of the flux losses for fake planets at the same radii\citep[e.g.,][]{2010A&A...509A..52C, 2013A&A...553A..60R}. The final $5\sigma$ contrast maps were obtained using the pixel-to-pixel noise maps divided by the flux loss and normalized by the relative calibration with the primary star, and were also corrected from small number statistics following the prescription of \citet{2014ApJ...792...97M} to adapt our $5\sigma$ confidence level at small angles. The azimuthal median of the contrast maps for all our targets with the sADI algorithm are reported in Fig. \ref{fig:contrast}.}

From these contrast maps and based on the COND model predictions \citep{2003A&A...402..701B}, we obtain detection mass maps, where the star's age, distance, and magnitude (reported in Table \ref{table:sample}) are considered for the luminosity-mass conversion.

\begin{figure*}[ht!]
\centering
\includegraphics[width=18cm]{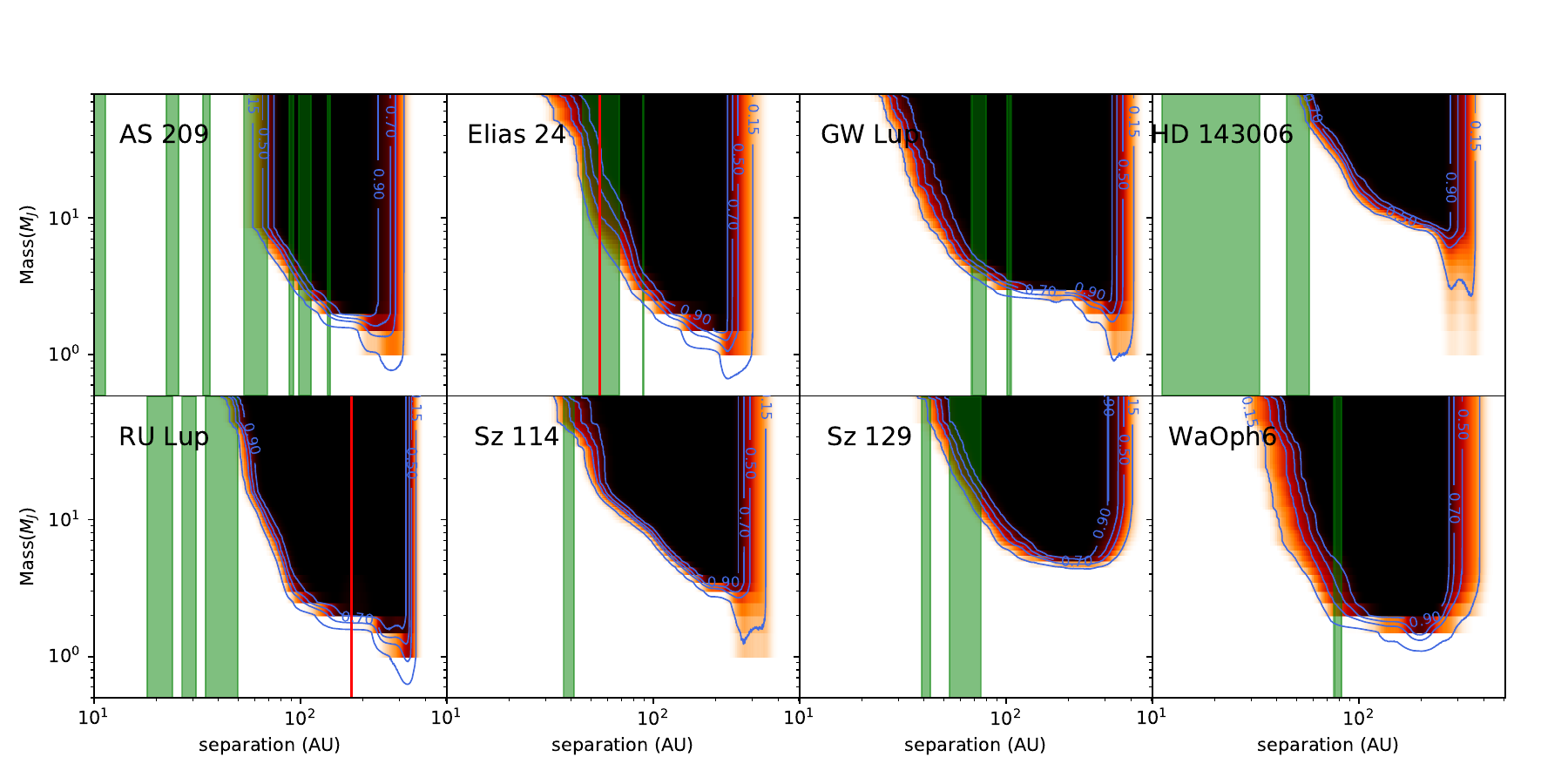}
\caption{Detection probability maps for the single-star targets in our survey. Shaded green regions correspond to the location and width of the observed gaps of the disks, as derived by \cite{2018ApJ...869L..42H}. Red lines indicate the location of the point sources presented in this work for Elias 24 and RU Lup. Our mass contrast limits are based on COND model predictions. In all cases, detection probability increases with both separation and mass, having a $50\%$ probability for masses $\sim 5M_J$ at $\approx100$ AU around all targets.}
\label{fig:mess}
\end{figure*}

Detection probability maps were then derived for 8 objects in our sample (excluding the known binaries HT\,Lup and AS\,205), by using the Multi-purpose Exoplanet Simulation System (MESS) code, a Monte Carlo tool for the predictions of exoplanet search results \citep{2012A&A...537A..67B}. We generated a uniform grid of mass and semi-major axis in the interval [0.5, 80]$M_J$ and [10, 1000] AU with a sampling of 0.5$M_J$ and 1 AU respectively. For each point in the grid, $10^4$ orbits were generated with a fixed inclination and position angle, $\Omega$ \citep[based on the inclination and position angle of the observed annular structures of the disk from ][]{2018ApJ...869L..42H}, but randomly oriented in space from uniform distributions in  $\omega, e < 0.1$ and $M$, which correspond to the argument of periastron with respect to the line of nodes, eccentricity, and mean anomaly, respectively. The detection probability maps are then built by counting the number of detected planets over the number of generated ones, by comparing the on-sky projected position (separation and position angle) of each synthetic planet with the 2D mass maps at $5\sigma$. The resulting detection probability maps are reported in Fig. \ref{fig:mess}, where the known gaps location is highlighted to demonstrate that for a fraction of the gaps we reach an interesting likelihood of massive planet detection.

\subsection{Point source detections}

\begin{figure*}[hb!]
\centering
\Large{\textsf{RU Lup}}

\includegraphics[width=12cm,trim=.4cm .6cm 0.5cm 1.5cm, clip]{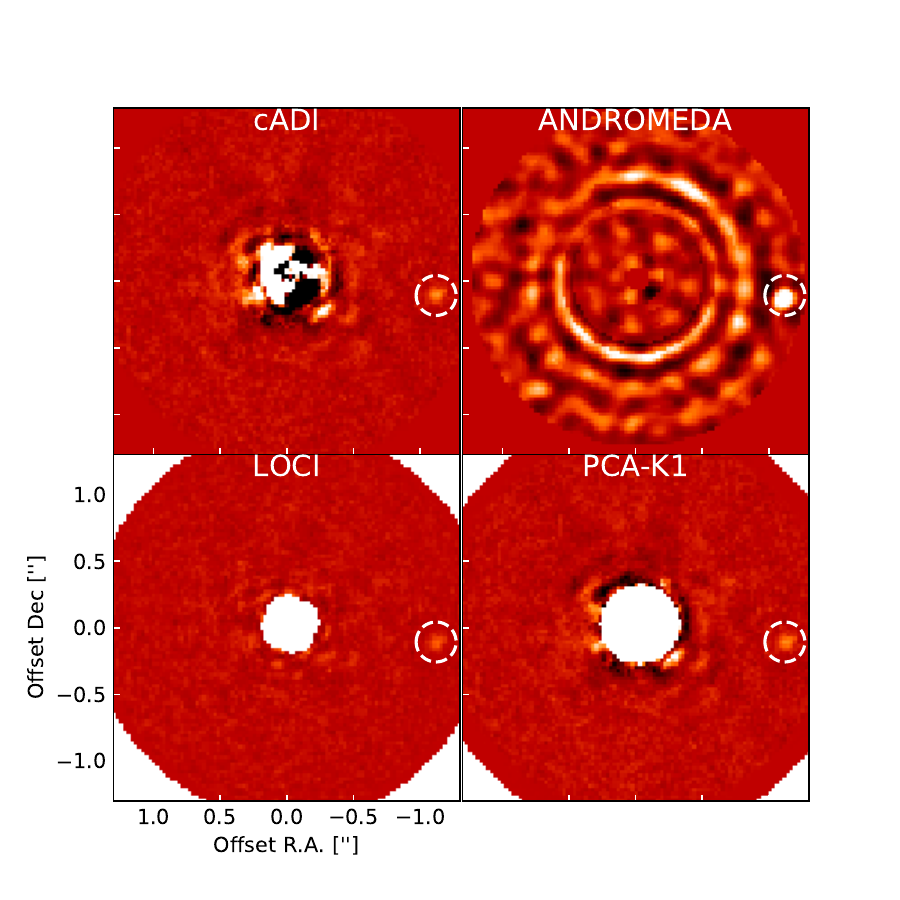}

\Large{\textsf{Elias 24}}

\includegraphics[width=12cm,trim=.4cm .6cm 0.4cm 1.5cm, clip]{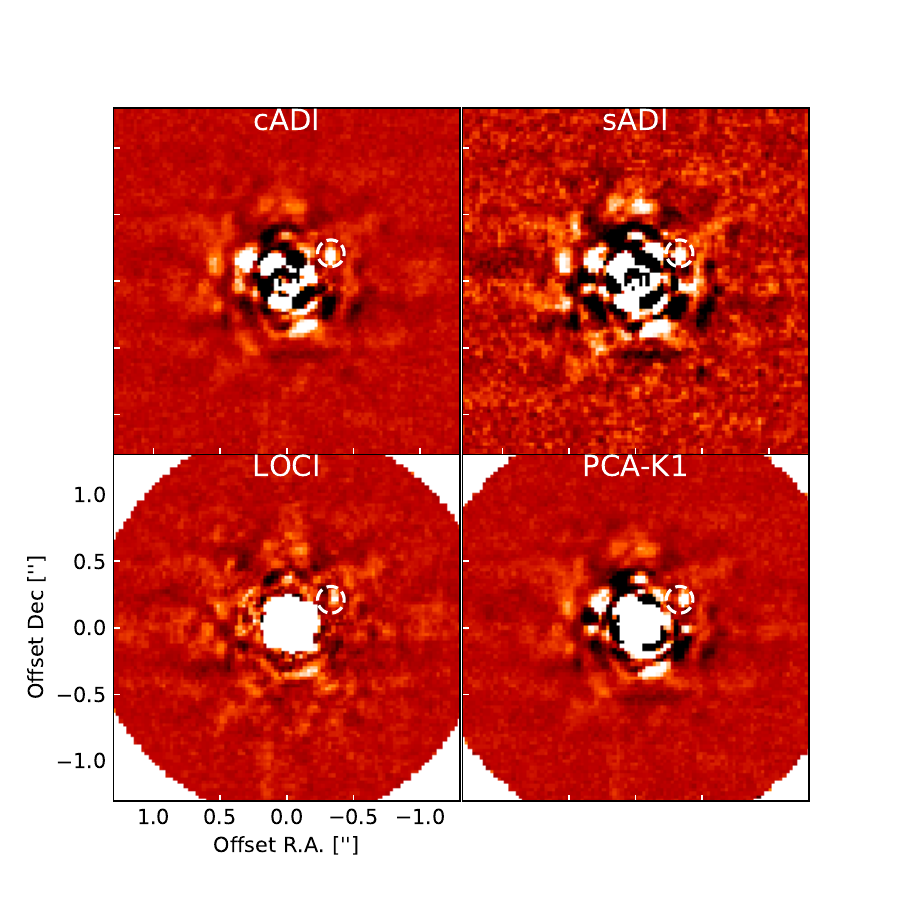}
\caption{L' band images for RU\,Lup (top) and Elias\,24 (bottom) after processing using four different ADI algorithms. A point source is recovered at the same position, marked with a circle, for all  algorithms.}
\label{fig:rulup adi}
\end{figure*}

Of the ten analysed systems, only RU Lup and Elias 24 exhibit a point like signal in our ADI reduction. Their signal appears consistently on all ADI processing methods, as shown in Fig. \ref{fig:rulup adi}, 
supporting the idea that we are in the presence of a real detection. Further analysis is then carried out for each object, aiming to characterize the properties of the possible companions.

\subsubsection{RU Lup}
The detected point source was first reported by \cite{2018ApJ...863...44A} using H band observations obtained with the SPHERE/IRDIS instrument on the VLT. It was regarded to be most likely a background object, although further discussion was not provided.

\begin{figure}[ht!]
\epsscale{1.2}
\plotone{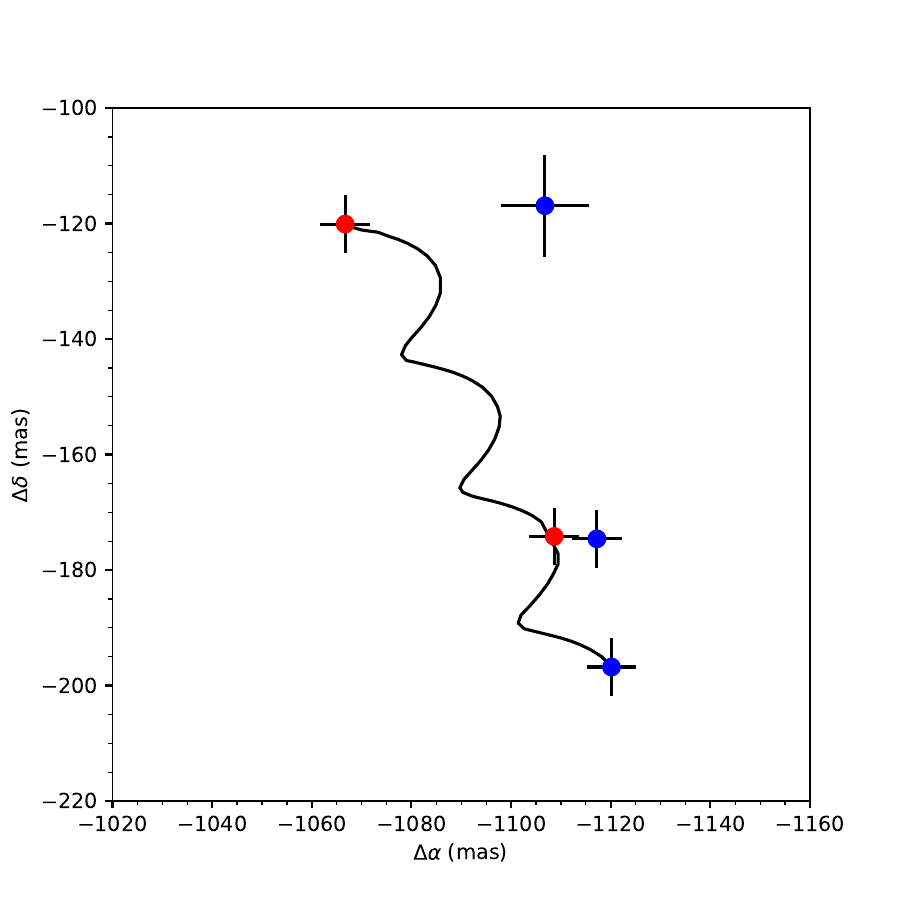}
\caption{Relative astrometry for the point-source detected in RU Lup, at three different epochs, from bottom to top: March-2016, March-2017 and May-2019. The predicted motion of a comoving object follows the black line, with the red dots being the expected astrometry of a comoving object at the different epochs of the observations. Blue dots are the real astrometry of the observed point-source at the different epochs of the observations. }
\label{fig:rulup astrometry}
\end{figure}

To characterize the position of the companion, negative "fake planets" were injected one by one  spanning a three dimensional parameter space (X position, Y position and flux) on the data cube before the PSF substraction. The cube was then reprocessed to derive the best-fit solution that minimize the residuals in the final subtracted image. This process located the companion at a separation of $1.112\pm0.008"$  and a position angle of $263.97\pm0.46^{\circ}$, relative to the host star of the system. In addition to the fake planet injection uncertainties, we considered additional errors coming from the central star position (saturated core, $<0.3$ pixels), and the platescale and True North calibration (see Section 3.). Using archival data from 2016 and 2017 (C. Ginski, priv. communication) we were able to derive the astrometry of the object. The results, displayed in Fig. \ref{fig:rulup astrometry}, indicate that the object is unambiguously not comoving with the primary star, with its astrometry being more consistent with that of a stationary background object. We still note a difference with the stationary background prediction, already present in the SPHERE measurements and confirmed with the NaCo third epoch, which suggests that the detected point-source has not a null proper motion.



\subsubsection{Elias 24}\label{subsec:elias}
In this case, the location of the point source is both extremely interesting and challenging. Following the same procedure as for the companion observed in the case of RU Lup, we locate the source at a separation of $0.411\pm0.008"$ ($\sim55$ AU) and a position angle of $302.1\pm1.14^{\circ}$. This result locates the source within one of the observed gaps of the disk, as can be seen in Fig. \ref{fig:sample}. It has been proposed that the annular structures in the disk formed due to the presence of a planetary companion at this location \citep{2017ApJ...851L..23C,2018MNRAS.475.5296D, 2018ApJ...869L..47Z}, supporting the possibility that the detected signal could be evidence of a planet forming in the disk. At the same time, our ADI observations starts degrading at close separations in the inner region of the disk, as evidenced by the presence of artifacts that can be observed at separations similar to the alleged signal. Because of this, multiple tests were carried out to determine the possibility of it being a true detection.

A first approach was to divide the original observing sequence into two, ensuring enough rotation of the field of view for the ADI algorithms to correctly process the images. In both sequences, the point-like signal was recovered at the same location for all the ADI methods used. A second test consisted on the injection of fake planets in the original data cube before the PSF subtraction. Fake planets were positioned at similar separations but different position angles ($0^{\circ}, 90^{\circ}$ and $180^{\circ}$) than the original signal. The cube was then processed using all ADI algorithms, to check if it was possible to recover the signal from the injected planets at the given separation. All the fake planets were recovered after the processing, producing similar point-like features as the original signal. The aforementioned tests strongly support the possibility that this detection could be of an actual planetary companion and not an speckle or other processing induced artifact. Another point to take into account is that first ADI detections of exoplanets using the NaCo instrument have been reported with a quality very similar to the ones obtained for Elias 24 \citep[e.g,][]{2009A&A...493L..21L,2018A&A...617A..44K}, indicating that it is still possible to recover the signal of a faint companion even at separation with a strong presence of residual artifacts after the data processing. 

To estimate the mass of the possible companion, we first derive its L' magnitude contrast based on the photometry of the signal, as well as the distance and L' magnitude of the host star, as reported on Table \ref{table:sample}. We derived the contrast to be $8.81\pm0.12$ mag. The age of the star is then considered in order to compare the measured magnitude contrast with the contrast expected for the appropriate isochrone of planetary evolution derived by \cite{2003A&A...402..701B}. This results in an estimated mass of $\sim 5M_J$ for the possible planet.

\begin{figure}[h!]
\centering
\includegraphics[width=9cm,trim=0cm 1cm 0cm 0cm, clip]{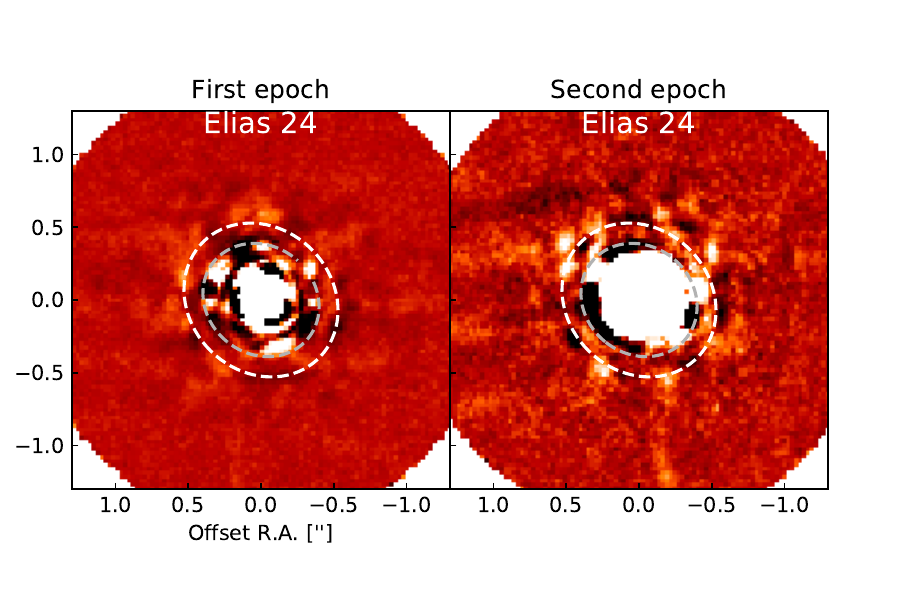}
\caption{Comparison of the two different epochs of observation of Elias 24, using the PCA method. The same color scale is used for both observations. The signal of the possible companion was not recovered again during the second epoch, which was reduced using the RDI method.}
\label{fig:elias epochs}
\end{figure}

Given our preliminary results, a new set observations on Elias 24 was carried out on September 2019, to confirm the preliminary signal. New observations could not be carried out in an identical setup, due to the closeness to the decommissioning date of NaCo and since the target was not optimal to be observed during transit. Thus, we opted for an RDI approach, as more relaxed observational constraints can be set to carry out these observations. Fig \ref{fig:elias epochs} shows the resulting image for the second epoch after RDI processing, compared with the cADI results from the first epoch. In this case the previously reported signal was not recovered. However, it has to be considered that the results obtained with the RDI processing are harder to interpret than the ADI results at the estimated location of the point-like source, given the numerous artifacts that remain after the data reduction, together with the fact that the gap is located very close to the inner limit of the region that can be resolved with RDI, thus difficulting the detection of a signal at that separation. Appendix B includes fake planet injection tests considering a bound companion or stationary background source hypothesis and showing that the second epoch is not conclusive. Based on this, further follow-up is mandatory to clarify the status of this faint planet candidate.

\section{Discussion}\label{sect:Discussion}
\subsection{Planet-disk interaction: comparison with models}
\begin{figure*}[ht!]
\centering
\includegraphics[width=18cm]{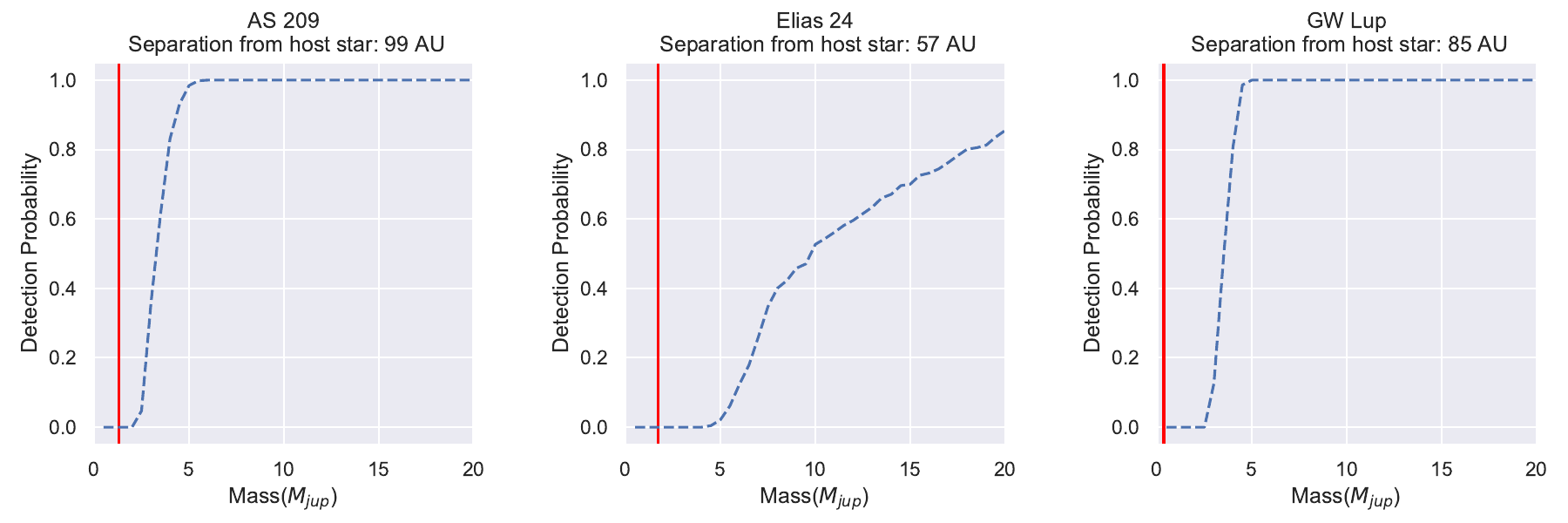}
\caption{Cut of our MESS Detection probabilities at the possible planet locations considered by \cite{2018ApJ...869L..47Z}.  Red vertical lines mark the highest mass needed for a planet to produce the observed gaps in the disk, as predicted by the hydrodynamical models.}
\label{fig:proba hydro}
\end{figure*}

\begin{figure*}[ht!]
\centering
\includegraphics[width=18cm]{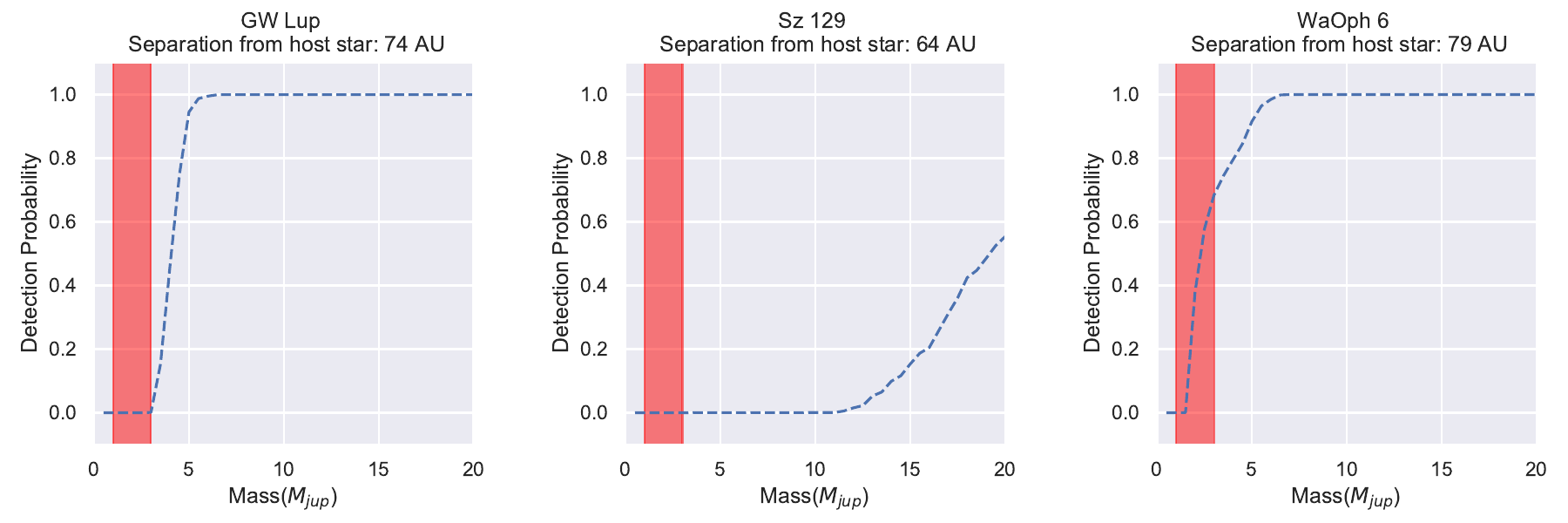}
\caption{Cut of our MESS Detection probabilities at the possible planet locations obtained by \cite{2020ApJ...890L...9P}. Red shaded intervals correspond to the mass range $1M_J - 3M_J$. In the three cases, planets location is consistent with the radius of one of the disk gaps.}
\label{fig:proba kine}
\end{figure*}

Assuming that the observed annular structures in our sample originate from the dynamical interaction between the disk and a forming planet, it is then necessary to analyze how our results, from both the detection and non-detection cases, correlate with the constraints on the system properties derived by different models of planet-disk interaction.

From Fig \ref{fig:contrast} and \ref{fig:mess}, one can observe that the separations at which detections are possible vary depending on each particular target, with the closest distance being between 30-80 AU for our full sample. However, at distances $>100\,$AU, we have a detection probability over $50\%$ for planets with masses around $5-10 M_J$ or higher, for almost all of our sample. Based on our non-detections, we can preliminary rule out the presence of high mass companions ($>5 M_J$) at these separations. 

Different studies have shown that both the disks properties and the mass of a planetary companion embedded are closely linked to the final morphology and substructures of the disk. For example, \citet{2018ApJ...864L..26B} found that a Jupiter-mass planet can produce different disk morphologies depending on the viscosity of the disk and its grain size distribution. Similarly, \citet{2018ApJ...869L..47Z} found from hydrodynamical simulations that companions with different masses can produce the structures observed on the disks of the DSHARP sample, depending on the disk surface density, viscosity and dust size distribution. A link between the companion properties and the morphology of the disk was also derived \cite{2019MNRAS.486..453L}, who inferred planetary masses of possible companions based on the width of the gaps observed in different protoplanetary disks observed by ALMA, which were consistent with the mass estimations derived from hydrodynamical models.

 Focusing on the results from \cite{2018ApJ...869L..47Z} there are three particular cases for which we can compare our MESS results for the derived detection probabilities with the planetary masses and disk properties estimated by these simulations: AS 209, Elias 24 and GW Lup. In Fig \ref{fig:proba hydro} we present our inferred detection probabilities for companions at the locations where hydrodynamical models require the presence of a planet to generate the observed gaps. For GW Lup,the highest estimated mass from hydro modeling for a companion is $0.06 M_J$. However, it is also mentioned that this value could increase up to $ 0.36 M_J$ if the gaps observed at 74 and 103 AU are part of a wide gap separated by a horseshoe region. In any case, the estimated mass is too low to be detected by our ADI observations. This is the same for AS 209, where the highest mass for a planet derived from the models is $1.32 M_J$. The case of Elias 24 is similar, as it is not possible to detect even the higher mass planet ($1.7 M_J$) predicted by the models. For a mass of $5M_J$, as the one derived in Sect \ref{subsec:elias} for the observed signal, we obtain a detection probability of $\sim 5\%$. Although it is still a low value, it might still be compatible with the one of a real planetary mass companion. For other targets, it was not possible to resolve the inner regions of the disk, so it is not possible to compare with theoretical predictions as no constraints in the companion masses can be derived at those close-in separations. However, the derived detection probabilities for all the disks in the sample do not take into account the possible presence of a circumplanetary disk (CPD), which might alter these values, although to an unknown extent. (see section 5.2 for further discussion)

 An alternative method to detect and characterize planetary companions in protoplanetary disks is to search for their kinematic signatures in the gas emission. Normally, for a given velocity channel, gas emission is concentrated along a Keplerian iso-velocity curve, corresponding to the region of the disk where the projected velocity is equal to the channel velocity and which is consistent with Keplerian rotation of the gas. However, the presence of a planet in the disk can distort the iso-velocity curve, producing a distinctive "kink" in the emission. Hydrodynamical models can then be used to reproduce the observed kinks considering an embedded planet in the disk, allowing for an estimation of the mass of the possible companion depending on the magnitude of the deviation observed. This technique has proven useful, as it has led to detections of kinks that may be associated with massive planets in the disks surrounding HD163296 \citep{2018ApJ...860L..13P} and HD97048 \citep{2019NatAs...3.1109P}. \cite{2020ApJ...890L...9P} reported the detection of kinematical signatures in nine disks from the DSHARP sample, using $^{12}$CO J=2-1 emission observations. 
 
 Four of these objects were also observed here: HD 143006, GW Lup, Sz 129 and WaOph 6. In all cases, the location of the velocity ``kink'' is consistent with the presence of a planet within a disk gap, suggesting that a significant fraction of the observed structures are caused by embedded protoplanets. Due to the velocity resolution of the observations, \cite{2020ApJ...890L...9P} cannot properly determine the mass of the possible perturber. Instead, ranges of masses from $1M_J$ to $3M_J$ were estimated for almost all cases, with the exception of HD 143006, whose velocity deviation is significantly larger, pointing to a more massive planet. Depending on the case,  \cite{2020ApJ...890L...9P} argues that mass estimations can range from 4 up to 10 times larger than the masses derived from the hydrodynamical models of \cite{2018ApJ...869L..47Z} and \cite{2019MNRAS.486..453L}, giving them a higher probability of being detected by direct imaging, based on Fig \ref{fig:mess}. In Fig \ref{fig:proba kine} we present our derived detection probabilities \emph{at the location of the gaps} for the cases of GW Lup, Sz 129 and WaOph 6 (HD 143006 was not included, as it is not possible to resolve the region where the planet is expected to lie, at a separation of 22 AU from the host star). For Sz 129, the estimated planet mass from the observed kinks is still too small to be detected in our ADI observations, as only planets with masses higher than $11 M_J$ are expected to be observable with any probability. The best options for a direct detection of a companion is then WaOph 6, which has a detection probability of $\sim70\%$ for planets with $3M_J$. GW Lup also remains as a good candidate for a direct detection, although for a slightly higher mass range, from $3M_J$ to $5M_J$. The fact that no signal of a possible companion was recovered after the ADI processing suggest the presence of planet with masses lower than $3M_J$. 
 
 The lack of detections of high mass planets at the location of the gap and even further away also gives important insight into the origin of the observed spiral arms in the disk of WaOph 6. \cite{2018ApJ...869L..43H} discussed the possibility of this spiral features to be created due to a planetary perturber located outside the arms, which extend up to 75 AU, and with a mass of several Jupiter masses. If that is the case, we should expect to recover the signal of these companions after the ADI processing based on the detection probabilities on Fig. \ref{fig:mess}, which reveal that we are sensitive to the detection of high mass planets at these regions. Although follow-up observations of this target are needed to definitely rule out the presence of a massive companion, as well as to check for possible planets with masses below our derived detection limits, this suggests that for WaOph 6, the spiral arms might not be triggered by interaction with a massive companion, and other alternatives should be considered such as formation due to gravitational instabilities. 

\subsection{Circumplanetary Disk effects}

 Taking into account the young age of the sample ($\leq10 \textrm{Myr}$), combined with the derived planet masses that are consistent with the formation of the observed annular substructures \citep{2018ApJ...869L..47Z}, it is expected that the planetary companions forming inside the DSHARP disks are on relatively early evolutionary stages, when protoplanets are still actively accreting. If that is the case, it is  expected that the accreted material will form a circumplanetary disk that will surround the planet \citep[CPD;][]{1998ApJ...508..707Q, 1999ApJ...526.1001L, 2005A&A...433..247P, 2009MNRAS.393...49A, 2012MNRAS.427.2597A, 2012A&A...547A.111M}. Any attempt to observe the forming protoplanets will then not only capture the emission from the planet, but from the CPD as well, making it crucial to address how the presence of a CPD impacts on both the detection and characterization of these systems. 


CPDs are expected to be very bright at infrared wavelengths, specially at the L'-band \citep{2015ApJ...803L...4E, 2015IAUGA..2256118Z, 2019MNRAS.487.1248S}, with its luminosity even surpassing that from the protoplanet. Because of this, the total emission of the CPD+planet system could mimic the one derived from evolutionary models for planets with masses up to an order of magnitude higher than the protoplanet real mass \citep{2019MNRAS.487.1248S}. Although this increased brightness can make these systems easier to detect, exact determination of their physical properties remains an extremely difficult challenge due to the multiple different phenomena CPDs are subjected to, such as the effects of magnetic fields \citep{2013ApJ...779...59G, 2014ApJ...783...14T, 2015MNRAS.451.1104K, 2016Sci...353.1519P}, accretion outbursts \citep{2012ApJ...749L..37L}, episodic infall of material from the protoplanetary disk \citep{2013ApJ...779...59G}, and radiative feedback from the protoplanet \citep{2015ApJ...806..253M, 2017arXiv171101372G}. Given that all of these processes play an important role on the final architecture and emission of the CPD, and since they can differ for each particular target, it is essential to study each system individually to properly characterize it.

A possible method to derive the properties of these systems is to use the relation of its apparent magnitude, which can be obtained from observations, with other properties such as its mass and accretion rate. It has been found that the brightness of these systems might be directly linked to the product of the planet mass and the mass accretion rate ($M_p\dot{M}$) as well as the CPD inner radius, $R_{in}$,  \citep[][]{2015ApJ...799...16Z, 2017A&A...608A..72M}. Thus, the contrast of a detected companion can be used to derive the value of $M_p\dot{M}$ for each particular system. This information can then be used to run system-specific disk models, aiming to reproduce both the signal of the companion and the structures observed on the protoplanetary disk. With these results, it is then possible to determine, or at least constrain, properties such as the mass or temperature of both the CPD and the embedded protoplanet for each particular system.

 
 The case of the point-source observed on the gap of Elias 24 can be used to illustrate the previously mentioned method. Based on the derived contrast of the signal, we estimate its apparent magnitude to be $L' = 15.4 \pm 0.14$. Assuming the distance of 136 pc reported in Table \ref{table:sample}, its absolute magnitude is $L' = 9.73 \pm 0.14$. This magnitude is then compared with the results from \citet{2015ApJ...799...16Z}, allowing to obtain three possible values of $M_p\dot{M}$ with an Absolute L' magnitude consistent with the value derived for the observed source. Finally, these values of $M_p\dot{M}$ are used to obtain an estimate of the accretion rates associated with the mass predictions from the hydrodynamical models of \citet{2018ApJ...869L..47Z} and the mass derived on Sect. \ref{subsec:elias}. The minimum mass is set at $0.4 M_J$, as derived from \citet{2018ApJ...869L..47Z}. Two values are considered for the maximum masses; $1.72 M_J$, which is the highest mass predicted by the hydrodynamical models, and $5 M_J$ which is the mass derived from evolutionary models based on the luminosity of the source. The predicted accretion rates are reported in Table \ref{tab:Elias}. For the ranges of masses and CPD inner radii we consider, the planet accretion rate inferred (from $1.2\times10^{-7}M_Jyr^{-1}$ up to $1.5\times10^{-5}M_Jyr^{-1}$) indicate a reasonable, low to moderate, growth for the posited planet. Future observations and dedicated hydro simulations are needed to confirm and further characterize the nature and the properties of the observed point-source in Elias 24.

\begin{deluxetable*}{ccccc}
\label{tab:Elias}
\tablenum{2}
\tablecaption{Planet-CPD parameters for Elias 24}
\tablewidth{0pt}
\tablehead{
\colhead{L'} & \colhead{$M_p\dot{M}$} & \colhead{$R_{in}$} & \colhead{Mass} & \colhead{Accretion rate} \\
\colhead{(mag)} & \colhead{($M_{J}^{2}\textrm{yr}^{-1}$)} & \colhead{($R_J$)} & \colhead{($M_J$)} & \colhead{$(M_{J}\textrm{yr}^{-1})$} 
}
\decimalcolnumbers
\startdata
9.8 & $6\times10^{-6}$ & 4 & 0.4 / 1.72 / 5& $1.5\times10^{-5}$ / $3.5\times10^{-6}$ / $1.2\times10^{-6}$  \\
9.8 & $10^{-6}$ & 2 & 0.4 / 1.72 / 5 & $2.5\times10^{-6}$ / $5.8\times10^{-7}$ / $2\times10^{-7}$  \\
9.5 & $6\times10^{-7}$ & 1 & 0.4 / 1.72 / 5& $1.5\times10^{-6}$ / $3.5\times10^{-7}$ / $1.2\times10^{-7}$ \\
\enddata
\tablecomments{Mass ranges and accretion rates based on the results from \citet{2015ApJ...799...16Z}. Given the uncertainties on the measurement of the L' magnitude of the point-like source on the gap of Elias 24, we also include the parameters  of the CPD model that produces the second closest L' magnitude to our derived value. The columns are ordered as follows: Column 1: Expected L' magnitude from the CPD models. Column 2: Expected value of $M_p\dot{M}$ from the CPD models. Columnn 3: CPD inner radius. Column 4: Companion mass range. Column 5: Accretion rates associated with the minimum and maximum mass of the planet, respectively.}
\end{deluxetable*}

The existence of a CPD can be inferred by other tracers than a potential overluminosity in infrared. A spectrum may actually reveal the presence of circumplanetary material significantly contributing to the total spectral energy distribution in addition to the planet's atmospheric emission as proposed for PDS70b combining SPHERE, NaCo and SINFONI observations at VLT \citep{2018A&A...617A..37C, 2018A&A...615A.160C, 2018A&A...617L...2M}. Shocks on the CPD surface can arise from meridional accretion flows \citep{2012ApJ...747...47T, 2014Icar..232..266M, 2014ApJ...782...65S, 2019ApJ...870...72D}, which happen at near free-fall speed and heat the gas to thousand of kelvin, theoretically producing strong $H_{\alpha}$ emission \citep{2018ApJ...866...84A}. $H_{\alpha}$ high-contrast imaging can then be used as an alternative technique to observe these systems. Surveys using this method have already been carried out with MagAO/VisAO, VLT/ZIMPOL and MUSE \citep[e.g; ][]{2014ApJ...781L..30C, 2018A&A...613L...5H, 2019A&A...622A.156C, 2019NatAs...3..749H, 2020A&A...633A.119Z} as well as observation on particular targets, producing interesting results such as a successful $H_\alpha$ detection at the same location as the imaged protoplanets PDS70b and c \citep{2018ApJ...863L...8W,2019NatAs...3..749H}. They definitively  offer a rich perspective to explore the physics of accretion (accretion rate, variability) during the formation of young giant planets in addition to their discoveries.

CPDs can be potentially detected also at longer wavelengths, such as the sub-millimetre regime. \citet{2018MNRAS.479.1850Z} found from multiple different models of CPDs that radio observations of these systems are more sensitive at shorter wavelengths, which in the case of ALMA are possible at Band 7 or above. Similarly, \citet{2018MNRAS.473.3573S} propose that the best option is to use Band 9 observations with the ALMA instrument to detect these systems, which allows to obtain fluxes up to 2 times higher than the flux detected in any other band. Furthermore, high resolution imaging at long wavelengths aiming to detect circumplanetary disks have already been carried out, with the most promising result being the detection of a third planetary companion orbiting the system PDS70 \citep[PDS70c; ][]{2019ApJ...879L..25I}. These results strongly support the use of radio observations as an alternative method to detect planetary companions.



\subsection{Extinction from protoplanetary disk material}

Finally, it is important to point out that the contrast values, and subsequently the detection probability maps, derived from our observations might be optimistic estimates, since we are not considering the effects of the disk material on the planets infrared emission. The presence of dust grains between the mid-plane of the disk, where the planet is expected to form, an its surface layer is expected to reduce the observed brightness of a source embedded into the disk, due to absorption and scattering from this solid material \citep{2015ApJ...807...64Q, 2017A&A...601A.134M}. 

\cite{2020MNRAS.492.3440S} derived the effect of extinction due to disk dust for multiple planet masses, separations from the host star, and IR bands. They found that for a $5M_J$ planet, extinction almost does not affect any IR band at any separation, since the planet is extremely efficient at removing the dust content along its orbit, opening a deep gap. On the other hand, it was found that for planets with masses below $5M_J$, dust extinction increases as the planet mass or the separation to the host star decreases, with dust extinction effects being less dramatic for longer IR wavelengths. Since the resolved regions after the ADI processing of our L' observations correspond to distances where dust extinction is supposed to be low ($\geq50 AU$), and that L-band is one of the bands less affected by extinction, we could expect that the derived detection limits will not vary much due to presence of dust near the planet.

\section{Conclusions}\label{sect:Conclusions}
In this work we present the first attempt to directly image planetary companions embedded into protoplanetary disks from the DSHARP sample, using angular differential imaging observations in thermal infrared. Our main results can be summarized as follows.
\begin{itemize}
    \item Observations of ten protoplanetary disks were processed using multiple flavours of ADI algorithms. The resolved regions vary for each disk, with a separation limit ranging from 30 to 80 AU, allowing for the detection of planets on the outermost parts of the disk. Possible detections of companions are reported for two disk: RU Lup and Elias 24.
    \item In the case of RU Lup, the possible companion is located at a separation of $\sim 1.1"$(174 AU) and has a estimated mass of $2.2 M_J$. Using archival data, the astrometry of the object was obtained for three different epochs, revealing it is not comoving with the host star of the disk. We conclude that it is in fact a stationary background object, being this the first time it is properly reported.
    \item In the case of Elias 24, the observed point-like signal is located within one of the observed gaps of the disk, at a separation of $\sim 0.42"$ (55 AU). Two mass ranges are derived for the companion, based on different assumptions. Evolutionary models suggest the mass to be at least $5M_J$, while a lower mass of 0.5 $M_J$ can be obtained when considering the presence of a CPD. Due to the challenging location of the signal, multiple tests were carried out to check for the possibility of it being an artifact induced by the reduction. Results from this tests seem to suggest that this is in fact a real detection.
    \item A second set of images of Elias 24 was obtained as a follow-up based on the possible detection previously reported. In this case observation could only be processed using an RDI technique. The point-like signal was not recovered this time.
    \item Contrast limits at $5\sigma$ were obtained in both magnitude and mass for each of the observed objects, as well as probability detection maps. These reveals that our observations are mostly sensitive to planets with masses $\geq5M_J$ at large separations ($\geq 100$ AU).
    \item Our detection limits and probabilities are compared with predictions on the companions masses based on both hydrodynamical simulations and kinematical signatures on the disk. In both cases the predicted masses are in good agreement with our results, since the putative companions would be too light and too faint to be detected in our survey. For WaOph 6, the lack of detections of high mass planets at the location of the gap or larger separations suggest that the observed spiral features on the disk are not originated due to the interaction of a planetary companion.
    \item Implications on the existence of an already formed CPD surrounding the theorized planetary companion in the gap of Elias 24 is discussed. Based on the derived luminosity of the point-like source and an inferred mass for the planet of $0.4 - 5 M_J$, we derived a range for the mass accretion rate of the system from $2\times10^{-8}M_Jyr^{-1}$ up to $1.5\times10^{-6}M_Jyr^{-1}$. In all studied cases this suggest a reasonable growth rate for the possible planet. Future observations and further modeling are needed to properly confirm if the observed signal is a real companion, if a CPD is surrounding the protoplanet, and to properly characterize this system.
\end{itemize}

Given our results, it is still too early to properly conclude if the observed structures on the disks are caused by the presence of planetary companions. Further attempts to observe these objects with more sensitive instruments, which will allow to go deeper in both contrast and location in the search of young planets, are currently in preparation. These multi-epochs observations are fundamental to determine the presence of a planet embedded into the disk via proper motion, and to derive its orbital properties, which will help to better understand the physical interaction between disks and planets, the evolution of protoplanetary disks, and the processes of planetary formation and evolution.

 \section*{Acknowledgments}

S.J.\ acknowledges support from the National Agency for Research and Development (ANID), Scholarship Program, Magister Becas Nacionales/2019 - 22191721.
L.M.P.\ acknowledges support from ANID project Basal AFB-170002 and from ANID FONDECYT Iniciaci\'on project \#11181068.
Support for J.H.\ was provided by NASA through the NASA Hubble Fellowship grant \#HST-HF2-51460.001-A awarded by the Space Telescope Science Institute, which is operated by the Association of Universities for Research in Astronomy, Inc., for NASA, under contract NAS5-26555. 
S.A.\ and J.H.\ acknowledge funding support from the National Aeronautics and Space Administration under Grant No.\ 17-XRP17 2-0012 issued through the Exoplanets Research Program.
J.M.C.\ acknowledges support from the National Aeronautics and Space Administration under grant No.\ 15XRP15\_20140 issued through the Exoplanets Research Program.
N.K.\ acknowledges support provided by the Alexander von Humboldt Foundation in the framework of the Sofja Kovalevskaja Award endowed by the Federal Ministry of Education and Research.
T.B.\ acknowledges funding from the European Research Council (ERC) under the European Union's Horizon 2020 research and innovation programme under grant agreement No 714769 and funding from the Deutsche Forschungsgemeinschaft under Ref. no. FOR 2634/1 and under Germany's Excellence Strategy (EXC-2094–390783311).

This paper makes use of the following ALMA data:
ADS/JAO.ALMA \#2016.1.00484.L, 
ADS/JAO.ALMA \#2011.0.00531.S, 
ADS/JAO.ALMA \#2012.1.00694.S, 
ADS/JAO.ALMA \#2013.1.00226.S, 
ADS/JAO.ALMA \#2013.1.00366.S, 
ADS/JAO.ALMA \#2013.1.00498.S, 
ADS/JAO.ALMA \#2013.1.00631.S, 
ADS/JAO.ALMA \#2013.1.00798.S,
ADS/JAO.ALMA \#2015.1.00486.S, 
ADS/JAO.ALMA \#2015.1.00964.S. 
ALMA is a partnership of ESO (representing its member states), NSF (USA) and NINS (Japan), together with NRC (Canada), MOST and ASIAA (Taiwan), and KASI (Republic of Korea), in cooperation with the Republic of Chile. The Joint ALMA Observatory is operated by ESO, AUI/NRAO and NAOJ. The National Radio Astronomy Observatory is a facility of the National Science Foundation operated under cooperative agreement by Associated Universities, Inc.

\vspace{5mm}
\facilities{VLT,ALMA.}


%




\appendix

\section{Obs log}

\begin{deluxetable*}{lccclccc}[h]
\label{table:obslog}
\tablecaption{Example observing log of targets with VLT/NaCo}
\tablewidth{0pt}
\tablehead{
\colhead{Type} & \colhead{Date} & \colhead{RA} & \colhead{DEC} & \colhead{Cam./filter} & \colhead{DIT $\times$ NDIT} & \colhead{$N_{\text{exp}}$} & \colhead{$\pi$-start/end}\\
& & \colhead{(hh:mm:ss)} & (dd:mm:ss) & & & &  }
\startdata
HD 143006 & 2018 Jul 12/13 & 15:58:36.91 & -22:57:15.22 & L27/L' & $0.2 \times 70$ & 280 &  $-99.1^{\circ}/99^{\circ}$ \\
HD 143006 PSF & 2018 Jul 12/13 & - & - & L27/L'+ND & $0.2 \times 50$ & 52 & - \\
AS 209 & 2018 Jul 13/14 & 16:49:15.30 & -14:22:08.26  & L27/L' & $0.2 \times 70$ & 280 & $-157.4^{\circ}/140.7^{\circ}$ \\
AS 209 PSF & 2018 Jul 13/14 & - & - &L27/L'+ND & $0.2 \times 50$ & 52 & -  \\
Elias 24 & 2018 Jul 15/16 & 16:26:24.09 & -24:16:13.46 & L27/L' & $0.2 \times 70$ & 260 &  $-94.8^{\circ}/95.3^{\circ}$ \\
Elias 24 PSF & 2018 Jul 15/16 & - & - & L27/L'+ND & $0.2 \times 50$ & 62 & - \\
WaOph 6 & 2019 Apr 28/29 & 16:48:45.63 & -14:16:35.85  & L27/L' & $0.2 \times 70$ & 300 &  $-129.8^{\circ}/122.5^{\circ}$ \\
WaOph 6 PSF & 2019 Apr 28/29 & - & - &L27/L'+ND & $0.2 \times 50$ & 51 & - \\
RU Lup & 2019 May 2/3 & 15:56:42.31 & -37:49:15.47  & L27/L' & $0.2 \times 70$ & 280 &  $-52.9^{\circ}/49.2^{\circ}$ \\
RU Lup PSF & 2019 May 2/3 & - & - & L27/L'+ND & $0.2 \times 50$ & 51 & - \\
GW Lup & 2019 Jun 3/4 & 15:46:44.73 & -34:30:35.68 & L27/L' & $0.2 \times 70$ & 188 &  $-61.6^{\circ}/13.4^{\circ}$ \\
GW Lup PSF & 2019 Jun 3/4 & - & - & L27/L'+ND & $0.2 \times 50$ & 31 & - \\
AS 205 & 2019 Jun 5/6 & 16:11:31.35 & -18:38:25.96  & L27/L' & $0.2 \times 70$ & 316 &  $-117.5^{\circ}/113.1^{\circ}$ \\
AS 205 PSF & 2019 Jun 5/6 & - & - & L27/L'+ND & $0.2 \times 50$ & 52 & - \\
Sz 129 & 2019 Jul 4/5 & 15:59:16.47 & -41:57:10.3  & L27/L' & $0.2 \times 70$ & 286 &  $-45.9^{\circ}/43.7^{\circ}$ \\
Sz 129 PSF & 2019 Jul 4/5 & - & - & L27/L'+ND & $0.2 \times 50$ & 52 & - \\
HT Lup & 2019 Sep 10/11 & 15:45:12.87 & -34:17:30.65 & L27/L' & $0.2 \times 70$ & 78 &  $87.6^{\circ}/91.9^{\circ}$ \\
HT Lup PSF & 2019 Sep 10/11 & - & - &  L27/L'+ND & $0.2 \times 50$ & 21 & - \\
Elias 24 & 2019 Sep 27/28 & 16:26:24.09 & -24:16:13.46 & L27/L' & $0.2 \times 70$ & 80 &  $99.9^{\circ}/101.6^{\circ}$ \\
2MASS J17354481-2413439 & 2019 Sep 27/28 & 17:35:44.82 & -24:13:43.88 & L27/L' & $0.2 \times 70$ & 80 & $98.8^{\circ}/100.5^{\circ}$ \\
Sz 114 & 2019 Sep 28/29 & 16:09:01.85 & -39:05:12.42 & L27/L' & $0.2 \times 70$ & 286 &  $-45.9^{\circ}/43.7^{\circ}$ \\
Sz 114 PSF & 2019 Sep 28/29 & - & - & L27/L'+ND & $0.2 \times 50$ & 52 & - 
\enddata
\tablecomments{ND refers to the NaCo ND\_long filter, PSF correspond to unsaturated exposures, DIT to exposure time and $\pi$ to the parallactic angle at start and end of the observation sequence}
\end{deluxetable*}

\section{Elias 24 second epoch observations}

\begin{figure}[h!]
\centering
\includegraphics[width=9cm,trim=0cm 1cm 0cm 0cm, clip]{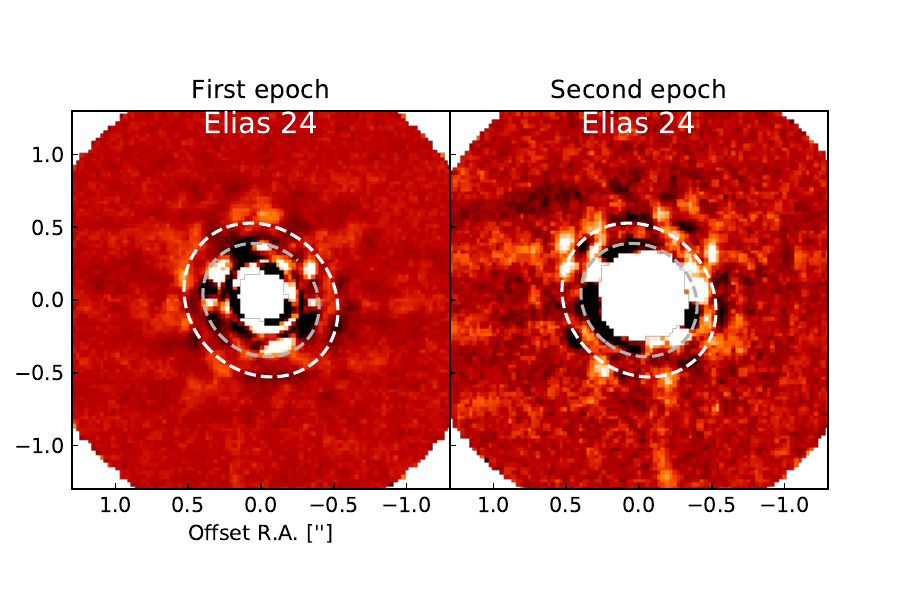}
\caption{Comparison of the two different epochs of observation of Elias 24, using the PCA method. A fake signal was injected at the expected location of the source observed during the first epoch, assuming it to be an stationary background object. Although the injected signal is slightly recovered, the poorer quality of the data is clear when compared to the first epoch.}
\label{fig:elias epochs bound}
\end{figure}

\begin{figure}[h!]
\centering
\includegraphics[width=9cm,trim=0cm 1cm 0cm 0cm, clip]{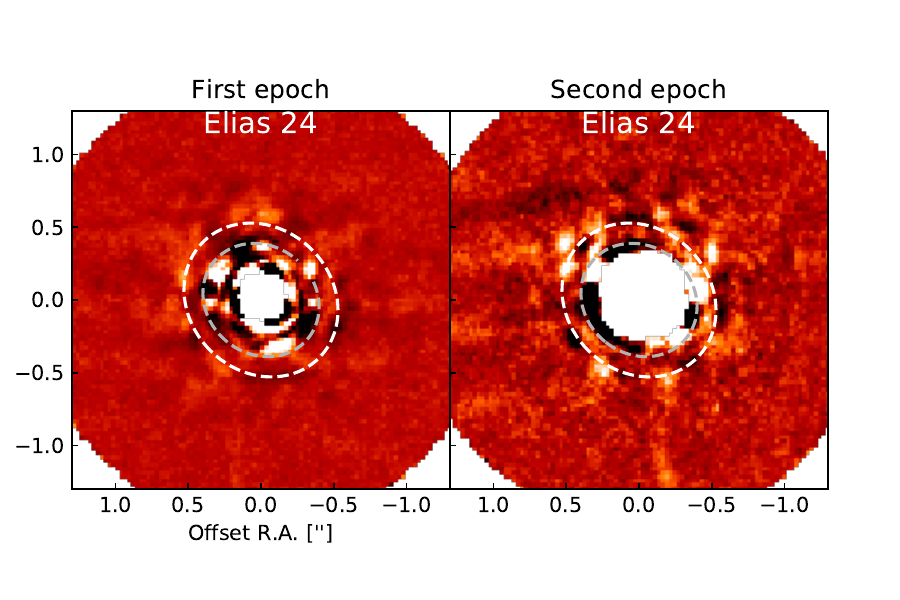}
\caption{Comparison of the two different epochs of observation of Elias 24, using the PCA method. A fake signal was injected at the expected location of the source observed during the first epoch, assuming it to be an bound companion. Although the injected signal is slightly recovered, the poorer quality of the data is clear when compared to the first epoch.}
\label{fig:elias epochs background}
\end{figure}

To test for the re-detection of the observed signal at the second epoch of observations for Elias 24, we performed a test consisting on the injection of a fake planet with the same flux as derived in the first epoch. The fake planet was injected both at a position assuming it to be a bound planet as well as the expected position of a background object. The results of the data reduction with this injected source are presented in Fig \ref{fig:elias epochs bound} and Fig \ref{fig:elias epochs background}. Given the small proper motion of the system \citep{2018yCat.1345....0G}, the location of the source appears to be the same in both cases.

From this injection test we can say that if the point source is real (bound planet or background source) the re-detection at the second epoch is extremely challenging given the poorer quality of the data. We advocate future observations with JWST, ERIS to clarify its nature (artefact, bound planet or background object).

\newpage

\bibliography{imaging}{}
\bibliographystyle{aasjournal}



\end{document}